\documentclass[twocolumn]{autart}    
\usepackage{graphicx}          
\usepackage{amsmath} 
\usepackage{amssymb}  
\usepackage{tikz} 
\usepackage{subfigure} 
\usepackage{enumerate} 
\usepackage{mathdots}

\begin{document}

\begin{frontmatter} 

\title{The Small Phase Condition is Necessary for Symmetric Systems\thanksref{footnoteinfo}} 

\thanks[footnoteinfo]{This work was supported in part by National Natural
Science Foundation of China under grants 62073003 and 72131001, by  the Natural Science Foundation of Guangdong Province under grants 2024A1515011630 and by the Research Grants Council of Hong Kong under the General Research Fund No. 16206324.   
}

\author[first]{Xiaokan Yang}\ead{yxkan21@stu.pku.edu.cn},    
\author[first,second]{Wei Chen}\ead{w.chen@pku.edu.cn},            
\author[third]{Li Qiu}\ead{qiuli@cuhk.edu.cn}  

\address[first]{Department of Mechanics and Engineering Science, Peking University, Beijing 100871, China}  
\address[second]{State Key Laboratory for Turbulence and Complex Systems, Peking University, Beijing 100871, China}             
\address[third]{School of Science and Engineering, The
Chinese University of Hong Kong, Shenzhen, Guangdong, China}    

\begin{keyword}                           
Small phase theorem; Feedback stability; Symmetric systems; Sectorial decompostion; Complex symmetric matrices.               
\end{keyword}                             

\begin{abstract}                 
In this paper, we show that the small phase condition is both sufficient and necessary to ensure the feedback stability when the interconnected systems are symmetric. Such symmetric systems arise in diverse applications. The key lies in that, for a complex symmetric and semi-sectorial matrix, the transformation matrix in its generalized sectorial decomposition can be taken to be real. Such a result fills the gap of phase based necessary condition for the feedback stability of symmetric systems, and serves as a counterpart of the necessity result for small gain condition. Moreover, we explore the necessity of small phase condition for general asymmetric systems. Some insightful results are presented, which help to clarify the main challenge in the general case. 
\end{abstract}

\end{frontmatter}

\section{Introduction} 

A new definition of phases of matrices and multi-input multi-output (MIMO) linear time-invariant (LTI) systems has been proposed recently \cite{Chen2024Phase,Mao2022Phases,Wang2020Phases,Wang2023Phases,Wang2024First_five_year,Furtado2003Spectral}, serving as a complementary pillar to the classical concept of gain. 
A small phase theorem, a counterpart of the small gain theorem, has been formulated and laid the foundation of a phase analysis framework for MIMO LTI systems \cite{Chen2024Phase,Mao2022Phases,Ringh2022GainPhaseMultipliers}. The phase analysis has already demonstrated its usefulness in various contexts. See for instance \cite{Wang2024Synchronization,Zhang2024Entangled,Chen2025Phase_N_port,Huang2024GainPhase,Fang2024Novel,Schmid2023MixedGainPhase,Ding2025Tensor}.

It is well known that the small gain condition is necessary if the uncertainty set is a ball of norm bounded systems. A natural question is whether the small phase condition is necessary. Some attempts have been made, such as extending the uncertainty set to include nonlinear systems \cite{Yang2023Small_phase_necessity,Yu2025Necessity_Small_Phase}, or considering the uncertainty to be a set of weighted stable positive real systems \cite{Chen2024Phase}. However, the necessity of small phase condition remains largely open. This paper aims to approach this gap. 

Our primary focus will be on symmetric systems, which occur in broad aspects and diverse applications \cite{Willems1972Dissipative_II,Willems1976Realization,Qiu1996Robustness,Yang2002SymmetricH2Controllers,Ikeda1993Optimality,Ikeda1995SymmetricControllers,Petersen2015PhysicalInterpretationNI,Petersen2010Feedback_control_NI}. For example, a space structure with collocated sensors and actuators may be described as 
\begin{equation*}
    M \ddot{q} + D \dot{q} + K q = Lu, \quad y = L^T q , 
\end{equation*} 
where the mass matrix $M$, damping matrix $D$, and stiffness matrix $K$ are all real symmetric. The transfer function matrix of this space structure, 
\begin{equation*}
    G(s) = L^T(Ms^2 + Ds + K)^{-1} L , 
\end{equation*} is clearly symmetric, i.e., $G(s) = G^T(s)$. Another important class of symmetric systems arises in circuit theory, namely reciprocal networks \cite{Anderson1973Network,Pates2022PassiveReciprocal} composed of resistors, inductors, capacitors and transformers (RLCT). We mention that all the single-input single-output (SISO) LTI systems are also included in the class of symmetric systems. 

In this paper, we will show that the small phase condition is not only sufficient but also necessary to ensure the feedback stability of symmetric systems. The key is that for a complex symmetric and semi-sectorial matrix, the transformation matrix in its generalized sectorial decomposition can be taken to be real. Moreover, we discuss the necessity of small phase condition in the general asymmetric case. The challenge is formalized as a phase interpolation problem, which remains open for future research. 

The remainder of this paper is organized as follows. In Section \ref{section:matrix_phase_factorization_complex_symmetric}, we introduce the notion of matrix phases and derive the factorizations of complex symmetric matrices. The small phase condition is shown to be both sufficient and necessary for the feedback stability of symmetric systems in Section \ref{section:Necessity_symmetric}. Some discussions on the necessity of small phase condition for asymmetric systems are given in Section \ref{section:discussions_general_LTI}. The paper is concluded in Section \ref{section:conclusion}. 

The notation used in this paper is mostly standard. Let $\mathbb{R}$ and $\mathbb{C}$ be the set of real and complex scalars. The conjugate, transpose and conjugate transpose of a matrix are denoted by $\overline{(\cdot)}$, $(\cdot)^T$ and $(\cdot)^H$, respectively. The Euclidean norm of a vector $x \in \mathbb{C}^n$ is defined as $\| x \| = \sqrt{x^Hx}$. The largest singular value of a matrix $C$ is denoted by $\overline{\sigma}(C)$. The boundary and interior of a set $S$ are denoted by $\partial S$ and $\mathrm{Int} S$, respectively. Let $\mathcal{R}^{m \times m}$ denote the set of $m\times m$ real rational proper transfer function matrices. The stable subset of $\mathcal{R}^{m \times m}$ is denoted by $\mathcal{RH}_\infty^{m \times m}$.

\section{Matrix Phases and Factorizations of Complex Symmetric Matrices} 
\label{section:matrix_phase_factorization_complex_symmetric}

\subsection{Phases of Matrices}   \label{subsection_phases_matrices}

We give a brief review on the phases of matrices. For more details, see \cite{Wang2020Phases,Wang2023Phases,Wang2024First_five_year,Furtado2003Spectral}. 
Given a matrix $C \in \mathbb{C}^{n \times n}$, its numerical range is defined as 
\begin{equation*}
     W(C) = \{ x^HCx: x \in \mathbb{C}^n, \| x \| = 1 \},  
\end{equation*} 
which is a compact and convex subset of $\mathbb{C}$ containing the spectrum of $C$ \cite{Horn1991Topics}. 
If $0\notin W(C)$, then $W(C)$ is contained in an open half complex plane and $C$ is said to be sectorial. Given a sectorial $C \in \mathbb{C}^{n \times n}$, there is a sectorial decomposition 
\begin{equation}                    \label{eq:sectorial_decomposition}
    C = T^HDT , 
\end{equation} 
where $T \in \mathbb{C}^{n \times n}$ is nonsingular and $D \in \mathbb{C}^{n \times n}$ is a diagonal unitary matrix \cite{Zhang2015Matrix}. The phases of $C$, denoted by 
\begin{equation*}
    \overline{\phi}(C) = \phi_1(C) \geq \phi_2(C) \geq \cdots \geq \phi_n(C) = \underline{\phi}(C) ,  
\end{equation*} 
are defined to be the phases of the eigenvalues of $D$ so that $\overline{\phi}(C) - \underline{\phi}(C) < \pi$. 
The phases defined in this way are not uniquely determined, but are rather determined modular $2\pi$. If one makes a selection of the phase center $\gamma(C)=\left[ \overline{\phi}(C)+\underline{\phi}(C) \right] / 2$ in $\mathbb{R}$, then the phases are uniquely determined. The phases are said to take principal values if $\gamma(C)$ is chosen in $[-\pi,\pi)$. We will not always select $\gamma(C)$ using its principal value. Following the standard way of selecting the phase of a complex scalar, we will select $\gamma(C)$ to make it continuous in the elements of $C$. In this way $\overline{\phi}(C)$ and $\underline{\phi}(C)$ are also continuous in the elements of $C$. 
Define the phase sector of $C$ as 
\begin{equation*} 
    \Phi(C) = [\underline{\phi}(C), \overline{\phi}(C)] . 
\end{equation*} 

\begin{figure}[htb]  
\centering
\begin{tikzpicture}
\begin{scope}
\draw [white, fill = gray, opacity = 0.5, rotate = -11.5](2.2,2.6) ellipse(0.5 and 1.25); 
\draw (0.8,1.7)--(3,3.5); 
\draw (0.8,1.7)--(3.3,0.45); 
\draw[->] (1.2,1.7) arc(0:39:0.4); 
\draw[->] (1.4,1.7) arc(0:-25:0.6); 
\draw [->] (1.55, 1.5) arc(10:-15:1);
\end{scope}
\draw[-latex, thick](0,1.7)--(3.8, 1.7); 
\draw[-latex, thick](0.8,0.4)--(0.8,3.8); 
\node[below] at (3.7,1.7){$\mathrm{Re}$}; 
\node[left] at(0.8,3.8){$\mathrm{Im}$}; 
\node[left] at(0.8,1.5){$0$}; 
\node[above] at(2.75,2.1){$W(C)$}; 
\node[right] at (1.3,1.95){$\overline{\phi}(C)$}; 
\node[below] at (1.7,1.15){$\underline{\phi}(C)$}; 
\end{tikzpicture} 
\vspace{-0.5em} 
\caption{The phases of a sectorial matrix.} 
\label{fig:sectorial_matrix}  
\end{figure}
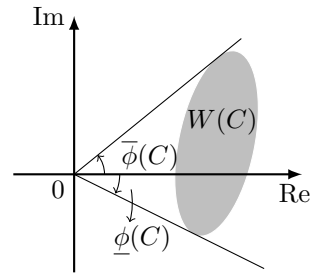

A geometric illustration of phases of a sectorial matrix is shown in Fig. \ref{fig:sectorial_matrix}. The two angles from the positive real axis to the two supporting rays of $W(C)$ are $\overline{\phi}(C)$ and $\underline{\phi}(C)$, respectively. The angle subtended by the two supporting rays is denoted by $\delta(C)$, and clearly $\delta(C) < \pi$.  

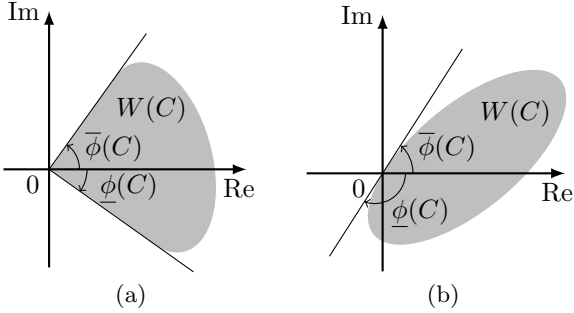
\begin{figure}[htb] 
\vspace{-0.5em} 
\centering 
\subfigure[]{
\begin{minipage}[t]{0.45\linewidth}
\centering 
\begin{tikzpicture}
\begin{scope}  
\draw [white, fill = gray, opacity = 0.5, rotate around = {15:(0.8,1.7)}](0.8,1.7)--(2,2.7) arc(110:-114:0.7 and 1.3)-- (0.8,1.7);  
\draw (0.8,1.7)--(2.08,3.5); 
\draw (0.8,1.7)--(2.7,0.34); 
\draw[->] (1.2,1.7) arc(0:54:0.4); 
\draw[->] (1.3,1.7) arc(0:-36:0.5);  
\end{scope} 
\draw[-latex, thick](0.2,1.7)--(3.4, 1.7); 
\draw[-latex, thick](0.8,0.4)--(0.8,3.8); 
\node[below] at (3.3,1.7){$\mathrm{Re}$}; 
\node[left] at(0.8,3.8){$\mathrm{Im}$}; 
\node[left] at(0.8,1.5){$0$}; 
\node[above] at(2.15,2.2){$W(C)$}; 
\node[right] at (1.15,2){$\overline{\phi}(C)$}; 
\node[below] at (1.85,1.77){$\underline{\phi}(C)$}; 
\end{tikzpicture} 
\label{fig:quasi_sectorial_matrix}
\end{minipage} 
}
\subfigure[]{
\begin{minipage}[t]{0.45\linewidth}
\centering 
\begin{tikzpicture}
\begin{scope}  
\fill [gray, opacity = 0.5, rotate = -50](0.02,3) ellipse(0.7 and 1.6); 
\draw (1.2,1.7)--(2.25,3.35); 
\draw (1.2,1.7)--(0.5,0.6); 
\draw[->] (1.6,1.7) arc(0:47:0.5); 
\draw[->] (1.5,1.7) arc(0:-110:0.4);  
\end{scope}
\draw[-latex, thick](0.2,1.7)--(3.6, 1.7); 
\draw[-latex, thick](1.2,0.4)--(1.2,3.8); 
\node[below] at (3.5,1.7){$\mathrm{Re}$}; 
\node[left] at(1.2,3.8){$\mathrm{Im}$}; 
\node[left] at(1.1,1.5){$0$}; 
\node[above] at(2.95,2.2){$W(C)$}; 
\node[right] at (1.55,2.05){$\overline{\phi}(C)$}; 
\node[below] at (1.7,1.5){$\underline{\phi}(C)$}; 
\end{tikzpicture} 
\label{fig:semi_sectorial_matrix}
\end{minipage} 
} 
\vspace{-0.5em} 
\caption{(a) The phases of a quasi-sectorial matrix; (b) The phases of a semi-sectorial matrix.} 
\end{figure}

The phase definition can be extended to a broader class of matrices. A matrix $C \in \mathbb{C}^{n\times n}$ is said to be quasi-sectorial if $\delta(C) < \pi$. A quasi-sectorial $C$ has a decomposition 
\begin{equation}                    \label{eq:quasi_and_sectorial_relation}
    C = U^H \begin{bmatrix}   
        0  &  0  \\  0  &  C_s 
    \end{bmatrix} U  , 
\end{equation} 
where $U$ is unitary, $C_s \in \mathbb{C}^{r \times r}$ is sectorial and $r = \mathrm{rank}(C)$, which means the range and kernel of $C$ are orthogonal and the compression of $C$ to its range is sectorial \cite{Furtado2003Spectral}. The phases of $C$ are then defined as the phases of $C_s$. A graphical illustration of a quasi-sectorial matrix and its phases is shown in Fig. \ref{fig:quasi_sectorial_matrix}. It can be seen that $0$ is a sharp point of the boundary of the numerical range. 

It is also possible that $0$ is on the smooth boundary of the numerical range. Such matrices fall into the class of semi-sectorial matrices. A matrix $C \in \mathbb{C}^{n \times n}$ is said to be semi-sectorial if $W(C)$ is contained in a closed half complex plane, i.e., $\delta(C) \leq \pi$. Such a matrix has a generalized sectorial decomposition \cite{Furtado2003Spectral} 
\begin{equation}                 \label{eq:generalized_sectorial_decomposition} 
    C = T^H \begin{bmatrix} 
        0_{n-r} & 0 & 0 \\ 
        0 & D & 0 \\ 
        0 & 0 & E 
    \end{bmatrix} T , 
\end{equation} 
where $T \in \mathbb{C}^{n \times n}$ is nonsingular, $ r = \mathrm{rank}(C), D = \mathrm{diag} \{ e^{j\phi_1}, \dots, e^{j\phi_m} \}$ with $\theta_0 + \frac{\pi}{2} \geq \phi_1 \geq \cdots \geq \phi_m \geq \theta_0 - \frac{\pi}{2},$ and 
\begin{equation} 
\label{eq:semi_sectorial_E_form} 
    E = e^{j\theta_0}  \, \mathrm{diag}  \left\{ 
     \begin{bmatrix}
     0  &  -j \\ -j  &  1   \end{bmatrix}, \cdots , \begin{bmatrix}
       0  &  -j \\ -j  &  1 \end{bmatrix} 
        \right\} , 
    \end{equation} 
where $\theta_0 \in \mathbb{R}$. In this case, the phases of $C$ are defined as $\phi_1, \dots, \phi_m$ and $\frac{r-m}{2}$ copies of $\theta_0 \pm \frac{\pi}{2}$. 
The numerical range and phases of a semi-sectorial matrix are illustrated in Fig. \ref{fig:semi_sectorial_matrix}. 

Recall that the singularity of matrix $I+AB$ plays an important role in the stability analysis of feedback systems. The following matrix small gain theorem, which shows that the small gain condition is both sufficient and necessary against norm bounded uncertainties, is well known. Let $\mathbb{F} \in \{ \mathbb{C}, \mathbb{R} \} $ denote a field. 

\begin{lem} [Matrix small gain theorem \cite{Stewart1990MatrixPerturbation}]            \label{lem:matrix_small_gain} 
    Let $A \in \mathbb{F}^{n \times n}$ and $\gamma > 0$. Then $I+AB$ is nonsingular for all $B \in \mathbb{F}^{n \times n}$ satisfying $\overline{\sigma}(B) \leq \gamma $ if and only if $\overline{\sigma}(A) < \frac{1}{\gamma}$. 
\end{lem}

The following matrix small phase theorem shows that the small phase condition is both sufficient and necessary against phase bounded uncertainties. 

\begin{lem} [Matrix small phase theorem \cite{Wang2023Phases}] 
\label{lem:matrix_small_phase} 
    Let $A \in \mathbb{C}^{n \times n}$ be quasi-sectorial and $0 \leq \beta - \alpha < 2\pi$. Then $I+AB$ is nonsingular for all semi-sectorial $B \in \mathbb{C}^{n \times n}$ satisfying $\Phi(B) \subset [\alpha, \beta]$ if and only if  
    \begin{equation*}
        \Phi(A) \subset (-\pi - \alpha, \pi - \beta) . 
    \end{equation*}
\end{lem}

\begin{rem}        \label{rem:real_matrix_small_phase_not_hold} 
    The fact that uncertainty $B$ is complex-valued is essential for the necessity result in Lemma \ref{lem:matrix_small_phase}, even when $A$ is real. If $B$ is restricted to be real, the necessity of the small phase condition no longer holds.  
\end{rem} 

\subsection{Factorizations of Complex Symmetric Matrices}  
\label{subsection:complex_symmetric_decomposition} 

This subsection is dedicated to factorizations of complex symmetric matrices, which are of central importance in the analysis of symmetric systems. A matrix $C \in \mathbb{C}^{n \times n}$ is said to be complex symmetric if $C = C^T$. 

\begin{lem} [Autonne–Takagi factorization \cite{Horn1990Matrix}] 
\label{lem:Autonne_Takagi_factorization} 
    Let $C \in \mathbb{C}^{n \times n}$. Then $C$ is complex symmetric if and only if there exists a unitary matrix $U \in \mathbb{C}^{n \times n}$ such that $C = U\Sigma U^T$, where $\Sigma$ is diagonal containing the singular values of $C$. 
\end{lem} 

Lemma \ref{lem:Autonne_Takagi_factorization} says that for a complex symmetric $C$, the unitary matrix $V$ in its singular value decomposition (SVD) $C = U\Sigma V^H$ can be taken as the conjugate of $U$, i.e., $V = \overline{U}$. 

In a similar spirit, we show that for a complex symmetric and semi-sectorial matrix $C$, the transformation matrix $T$ in its generalized sectorial decomposition \eqref{eq:generalized_sectorial_decomposition} can be chosen to be real. This property is critical in subsequent treatment of symmetric systems. 

\begin{thm}             \label{thm:complex_symmetric_semi_matrix_decomposition} 
    Let $C \in \mathbb{C}^{n \times n}$. Then $C$ is complex symmetric and semi-sectorial if and only if there exists a nonsingular matrix $T \in \mathbb{R}^{n \times n}$ such that 
    \begin{equation}                \label{eq:semi_sectorial_symmetric_decomposition} 
        C = T^T  \begin{bmatrix}
            0_{n-r} & 0  &  0 \\ 0 &   D  & 0  \\  
            0  &  0  &  E 
        \end{bmatrix} T , 
    \end{equation} where $r = \mathrm{rank}(C)$, $D = \mathrm{diag} \{ e^{j\phi_1}, \dots, e^{j\phi_m} \}$ with $\theta_0 + \frac{\pi}{2} \geq \phi_1 \geq \cdots \geq \phi_m \geq \theta_0 - \frac{\pi}{2}$ and $E$ is given in \eqref{eq:semi_sectorial_E_form} where $\theta_0 \in \mathbb{R}$. 
\end{thm} 

Before proving Theorem \ref{thm:complex_symmetric_semi_matrix_decomposition}, we introduce some useful results. A canonical reduced form for a pair of real symmetric matrices under simultaneous real congruence transformation was established in \cite{Thompson1991Pencils}. 
Applying \cite[Theorem 2]{Thompson1991Pencils} to a complex symmetric matrix $C$, whose real part $C_R=\frac{1}{2}(C + \overline{C}) $ and imaginary part $ C_I = \frac{1}{2j}(C - \overline{C})$ are both real symmetric, one can infer that there exists a real congruence transformation that reduces $C$ to a direct sum of blocks of the following four types: $\delta K_k, \delta L_l(a), M_{2m-1}$ and $N_{2n}$, where $\delta \in \{ -1,1 \}, $ and $K_k \in \mathbb{C}^{k\times k}, L_l(a) \in \mathbb{C}^{l \times l}, M_{2m-1} \in \mathbb{C}^{(2m-1)\times(2m-1)}, N_{2n} \in \mathbb{C}^{2n \times 2n}$ are defined as  
\begin{align*} 
        K_k & = \begin{bmatrix} 
            0  &  \  &  \ &  1 \\ 
            \  &  \  &  \iddots & -j \\ 
            \  &  \iddots &  \iddots & \ \\ 
            1  & \  -j  & \  &  0 
        \end{bmatrix}  , 
        \end{align*} 
        
        \begin{align*}   
        L_l (a) & = \begin{bmatrix}
            0  &  \  &  \ &  a-j \\ 
            \  &  \  &  \iddots & 1 \\ 
            \  &  \iddots &  \iddots & \ \\ 
            a-j  &  1  & \  &  0  
        \end{bmatrix} ,  \\ 
        M_{2m-1} & = \begin{bmatrix}
            0_{m} & J \\ J^T &  0_{m-1} 
        \end{bmatrix} \text{ with } 
        J = \begin{bmatrix}
            -j & \  & \   &  0  \\ 
            1  & \  \ddots  & \!  & \\ 
            \  &  \ddots  & \  & -j \\ 
            0  &  &  &  1 
        \end{bmatrix}  ,  \\ 
        N_{2n}  & = \begin{bmatrix} 
            0  &  \  &  \ &  R \\ 
            \  &  \  &  \iddots & S \\ 
            \  &  \iddots &  \iddots & \ \\ 
            R  &  S  & \  &  0 
        \end{bmatrix} \text{ with }   
        \left\{  
        \begin{array}{l} 
             R =  \begin{bmatrix}
            b & a - j \\  a - j &  -b 
        \end{bmatrix}   \\
             S = \begin{bmatrix}
                0  &  1  \\  1   &   0  
            \end{bmatrix} 
        \end{array} 
        \right.  , 
    \end{align*} 
with $a,b \in \mathbb{R}, b \neq 0$ and $  K_1 = 1,L_1(a) =a-j,M_{1} =0,N_2=R $. 
The following lemma is useful in proving Theorem \ref{thm:complex_symmetric_semi_matrix_decomposition}. 

\begin{lem}             \label{lem:general_congruent_form}
    Let $a,b$ be real numbers and $k,l,m,n$ be positive integers. Then the following statements are true. 
    \begin{enumerate}[(i)]
        \item $0 \in \partial W(K_2)$ and $0 \in \mathrm{Int} W(K_k)$ for $k \geq 3$; 
        \item $0 \in \partial W(L_2(a))$ and $0 \in \mathrm{Int}W(L_l(a))$ for $l \geq 3$; 
        \item $0 \in \mathrm{Int} W(M_{2m-1})$ for $m \geq 2$; 
        \item If $b \neq 0$, then $0 \in \mathrm{Int} W(N_{2n})$ for $n \geq 1$. 
    \end{enumerate}
\end{lem} 

\begin{pf}
     The proof is inspired by the argument in \cite[Lemma 3]{Furtado2003Spectral}. 
     To show (i) and (ii), note that for $k \geq 2, l \geq 2$, $K_k $ and $ L_l$ have a principal entry equal to $0$, implying $0 \in W(K_k)$ and $0 \in W(L_l(a))$. A further calculation shows that $0 \in \partial W(K_2) $ and $ 0 \in \partial W(L_2(a))$. For $k,l \in \{ 3, 4\}$, one can verify that $0 \in \mathrm{Int} W(K_k)$ and $0 \in \mathrm{Int}W(L_l(a))$. When $k\geq 3$ ($l\geq 3$, respectively) is odd, then $K_3$ ($L_3(a)$, respectively) is a principal submatrix of $K_k$ ($L_l(a)$, respectively). When $k \geq 4$ ($l \geq 4$, respectively) is even, then $K_4$ ($L_4(a)$, respectively) is a principal submatrix of $K_k$ ($L_l(a)$, respectively). This yields that (i) and (ii) hold. To show (iii), note that $0 \in W(M_{2m-1})$ for $m \geq 2$ since $M_{2m-1}$ has a principal entry equal to $0$. A calculation shows that $0$ is actually in the interior of $W(M_{2m-1})$, implying (iii) holds. To show (iv), suppose $b \neq 0$, then $0 \in \mathrm{Int}W(N_{2n})$ for $n \in \{1, 2\}$. 
     When $n \geq 3$ is odd, then $N_2$ is a principal submatrix of $N_{2n}$. When $n \geq 4$ is even, then $N_4$ is a principal submatrix of $N_{2n}$. Hence, (iv) holds. This completes the proof. 
\end{pf}

With the above preparation, we are now in a position to prove Theorem \ref{thm:complex_symmetric_semi_matrix_decomposition}.  

\begin{pf}[Proof of Theorem \ref{thm:complex_symmetric_semi_matrix_decomposition}] 
    Sufficiency: In view that $E = E^T,D=D^T$ and $C$ has the form \eqref{eq:semi_sectorial_symmetric_decomposition}, one can verify that $C = C^T$. Since $0$ is not contained in the interior of the numerical range of $\mathrm{diag}\{  0_{n-r}, D, E\}$, it follows that $0$ is also not contained in the interior of $ W(C)$, implying $C$ is semi-sectorial. \\ 
    Necessity: As established in the preceding analysis, $C$ is real congruent to a direct sum of blocks of the following types: $\delta K_k, \delta L_l(a), M_{2m-1}$ and $N_{2n}$, with $\delta \in \{ -1,1 \}, a, b \in \mathbb{R}, b \neq 0$ and $K_k,L_l(a),M_{2m-1}, N_{2n}$ defined earlier. Since $C$ is semi-sectorial, it follows from Lemma \ref{lem:general_congruent_form} that $k \leq 2, l \leq 2, m \leq 1$ and no blocks of type $N_{2n}$ can appear;  otherwise, $0 \in \mathrm{Int} W(C)$. Hence $C$ is real congruent to a matrix of the form $Q_1 \oplus Q_2$, where $Q_1$ is diagonal and $Q_2$ is a direct sum of $2$-by-$2$ blocks. Moreover, $W(Q_1 \oplus Q_2)$ is contained in a closed half complex plane, implying $Q_1 = \begin{bmatrix}
          0_{n-r} & 0 \\ 0  &  D 
      \end{bmatrix}$, and only one of the blocks $K_2,-K_2,L_2(a),-L_2(a)$ can appear in $Q_2$.  \\ 
      Denote $ \tilde{E} = \begin{bmatrix}
          0  &  -j \\ -j  &  1
      \end{bmatrix} $ and 
    \begin{equation*}
         T_K = \begin{bmatrix}
          1 & 0 \\ 0 &  -1 
      \end{bmatrix} , \quad T_L(a) = \frac{1}{\sqrt{a^2+1}} \begin{bmatrix}
              a^2 \! + \! 1 &  \frac{a}{2} \\ 0 &  1 
          \end{bmatrix}  . 
    \end{equation*} 
    Then $K_2$ and $L_2(a)$ can be decomposed as 
      \begin{align*}
          K_2 & = \begin{bmatrix}
              0 & 1 \\  1  &  -j 
          \end{bmatrix} = -j T_K^T\tilde{E} T_K,  \\ 
          L_2(a) & = \begin{bmatrix}
              0 & a - j \\  a - j & 1 
          \end{bmatrix} = (1+aj) T_L^T(a) \tilde{E} T_L(a)  . 
      \end{align*} 
       This implies that there exists $\theta_0 \in \mathbb{R}$ such that $Q_2$ is real congruent to the direct sum of $e^{j\theta_0} \tilde{E}$. Hence there exists real matrix $T$ such that $C$ has the form \eqref{eq:semi_sectorial_symmetric_decomposition}. This completes the proof. 
\end{pf}

\begin{exmp} 
    Consider the matrix $\begin{bmatrix} 
        1  &  2  \\  0  &  1 
    \end{bmatrix}$. It is semi-sectorial and admits a generalized sectorial decomposition via a complex congruence transformation 
    \begin{equation*}
        \begin{bmatrix}
            1  &  2 \\ 0  &  1 
        \end{bmatrix} = 
        \begin{bmatrix}
            1  &  0  \\ j  &  j 
        \end{bmatrix}^H \begin{bmatrix}
            0  &  -j  \\ -j  &  1  
        \end{bmatrix} 
        \begin{bmatrix}
            1  &  0  \\ j  &  j 
        \end{bmatrix} . 
    \end{equation*} 
    However, since it is not complex symmetric, it does not admit a generalized sectorial decomposition via a real congruence transformation. 
\end{exmp}

The following corollary specializes the statement in Theorem \ref{thm:complex_symmetric_semi_matrix_decomposition} to the case of quasi-sectorial matrices. The proof is omitted for brevity. 

\begin{cor}             \label{cor:complex_symmetric_quasi_matrix_decomposition} 
    Let $C \in \mathbb{C}^{n \times n}$. Then $C$ is complex symmetric and quasi-sectorial if and only if there exists a nonsingular matrix $T \in \mathbb{R}^{n \times n}$ such that 
    \begin{equation*} 
        C = T^T  \begin{bmatrix}
            0_{n-r} & 0 \\ 0 &   D
        \end{bmatrix} T , 
    \end{equation*} where $r = \mathrm{rank}(C)$, $D = \mathrm{diag} \{ e^{j\phi_1}, \dots, e^{j\phi_r} \}$ with $\theta_0 + \frac{\pi}{2} >\phi_1 \geq \cdots \geq \phi_r > \theta_0 - \frac{\pi}{2}$ where  $\theta_0 \in \mathbb{R}$. 
\end{cor}

\section{Necessity of Small Phase Condition for Symmetric Systems}
\label{section:Necessity_symmetric}

For a system $G \in \mathcal{RH}_\infty^{m \times m}$, the singular values of $G(j\omega)$ are continuous functions of the frequency, which are called gain response of $G$. The $\mathcal{H}_\infty$ norm of $G$ is defined as \cite{Zhou1996Robust} 
\begin{equation*}
    \| G \|_\infty = \sup_{\omega \in [0,\infty]} \overline{\sigma} (G(j\omega)) . 
\end{equation*} While gain is defined only for stable systems, the phase notion can be defined for semi-stable systems with possible poles on the imaginary axis. In this paper, we focus on Lyapunov stable systems, i.e., systems whose poles lie in the closed left half plane with all the imaginary axis poles semi-simple. 

Let $G \in \mathcal{R}^{m \times m}$ be Lyapunov stable and denote the set of its imaginary axis poles by $j\Omega$. Then $G$ is said to be frequency-wise semi-sectorial if \cite{Wang2024First_five_year}
\begin{enumerate}[1)] 
    \item $G(j\omega)$ is semi-sectorial for all $\omega \in [-\infty,\infty] \setminus \Omega$; 
    \item For any $j\omega_0 \in j \Omega$, the residue matrix $\lim_{s \to j\omega_0} (s-j\omega_0) G(s)$ is semi-sectorial. 
\end{enumerate} 
The frequency-wise quasi-sectorial (sectorial, respectively) systems can be defined similarly.  

For a Lyapunov stable frequency-wise semi-sectorial system $G$,  if $0$ is neither a pole nor a zero of $G$, we assume for simplicity that $G(0)$ is accretive, i.e., $\mathrm{Re}(G(0)) \geq 0$, such that $\gamma (G(0)) = 0$. If $0$ is a pole or a zero of $G$, we look at $G(\epsilon)$ instead and assume that $G(\epsilon)$ is accretive. For both cases, let $\gamma(G(s))$ be defined continuously along the indented imaginary axis where right half-circle detours with radius $\epsilon$ are taken at both the poles and finite zeros of $G(s)$ on the imaginary axis and a right half-circle detour with radius $1 / \epsilon$ is taken if $\infty$ is a zero of $G(s)$. In this way, $\left[ \underline{\phi} (G(s)), \overline{\phi} (G(s)) \right]$ is a continuous interval-valued function on the indented imaginary axis, which is called the phase response of $G$. The phase response is odd with respect to the indented imaginary axis and thus it suffices to consider its part over the upper half of the indented imaginary axis. The largest and smallest phases of $G$ are defined as \cite{Wang2024First_five_year}
\begin{align*} 
\overline{\phi}(G) \! =\! \!  \sup_{\omega \in [0, \infty] \setminus \Omega } \! \overline{\phi}(G(j\omega)), \ \ 
\underline{\phi}(G) \! = \! \! \inf_{\omega \in [0,\infty] \setminus \Omega }  \underline{\phi}(G(j\omega))  , 
\end{align*} 
respectively. The $\Phi_\infty$ sector of $G$ is defined as
\begin{equation*} 
    \Phi_\infty (G) = [\underline{\phi}(G), \overline{\phi}(G)] , 
\end{equation*} 
which serves as a phasic counterpart of the $\mathcal{H}_\infty$ norm.

\begin{figure}[htb] 
    \setlength{\unitlength}{1mm}
		\begin{center}
			\begin{picture}(66,32)
				\thicklines 
				\put(13,6){\vector(0,1){18}} 
                \put(13,26){\circle{4}} 
                \put(0,26){\vector(1,0){11}}
				\put(15,26){\vector(1,0){13}} 
				\put(28,21){\framebox(10, 10){$G$}} 
				\put(38,26){\line(1,0){15}} 
				\put(53,26){\vector(0,-1){18}} 
				\put(53,6){\circle{4}} 
				\put(66,6){\vector(-1,0){11}}
				\put(51,6){\vector(-1,0){13}} 
				\put(28, 1){\framebox(10,10){$H$}} 
				\put(28,6){\line(-1,0){15}}  
				\put(18,26){\makebox(5,5){$u_1$}} 
				\put(42,6){\makebox(5,5){$u_2$}} 
				\put(14,18){\makebox(5,5){$-$}} 
				\put(59,6){\makebox(5,5){$e_2$}}
                \put(2,26){\makebox(5,5){$e_1$}} 
			\end{picture}
			\vspace{-0.5em} 
			\caption{A feedback system $G \# H$. } 
			\label{fig:feedback_system}
		\end{center}
\end{figure}
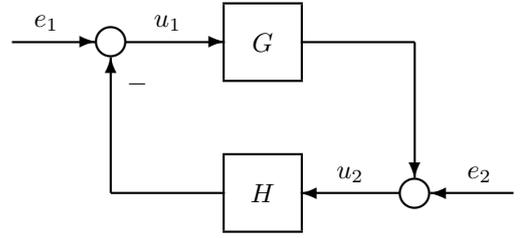

Now, consider the feedback interconnection of $G, H \in \mathcal{R}^{m \times m}$ as depicted in Fig. \ref{fig:feedback_system}. It is said to be stable if  
\begin{equation*}
    G\#H = \begin{bmatrix}
        I - H(I + GH)^{-1}G & \  - H(I + GH)^{-1} \\ 
        (I+GH)^{-1}G & (I+GH)^{-1}  
    \end{bmatrix} 
\end{equation*} 
is stable, i.e., $G \# H \in \mathcal{RH}_\infty^{2m \times 2m} $. 

The celebrated small gain theorem certifies the stability of the feedback system based on the gains of loop components.

\begin{lem}[Small gain theorem \cite{Zhou1996Robust,Liu2016Robust_Control}] 
    \label{lem:small_gain}
    Let $G,H\in \mathcal{RH}_\infty^{m \times m}$, then $G\#H$ is stable if 
    \begin{equation*}
        \overline{\sigma}(G(j\omega)) \overline{\sigma}(H(j\omega)) < 1 , 
    \end{equation*} 
    for all $\omega \in [0, \infty]$. 
\end{lem} 

As a counterpart, the recently established small phase theorem certifies the stability of the feedback system based on the phases of loop components.

\begin{lem}[Small phase theorem \cite{Wang2024First_five_year}]   
    \label{lem:small_phase} 
    \! Let $G \!\! \in \!\!  \mathcal{R}^{m \times m} $ be frequency-wise quasi-sectorial and $ H \in \mathcal{RH}_\infty^{m \times m}$ be frequency-wise semi-sectorial. Denote by $j\Omega$ the set of imaginary axis poles of $G$. Then $G\#H$ is stable if 
    \begin{align*}
        \overline{\phi}(G(j\omega)) + \overline{\phi}(H(j\omega)) & < \pi, \\  
        \underline{\phi}(G(j\omega)) + \underline{\phi}(H(j\omega)) & > - \pi , 
    \end{align*} 
    for all $\omega \in [0, \infty] \setminus \Omega$. 
\end{lem} 

It is known that the small gain condition is necessary in the following sense. Let $\gamma: \mathbb{R}\cup \{ \infty \} \to \mathbb{R}$ be a scalar function and define a ball of systems 
\begin{equation*}
    \mathcal{B}[\gamma] \! = \! \{ H \! \in \! \mathcal{RH}_\infty^{m \times m}: \overline{\sigma} (H(j\omega)) \! \leq \! | \gamma(\omega) |, \forall \omega \! \in \!  [0,\infty] \} . 
\end{equation*} 
Then $G\#H$ is stable for all $H \in \mathcal{B}[\gamma]$ if and only if 
$\overline{\sigma}(G(j\omega)) < |\gamma(\omega)|^{-1} $ for all $\omega \in [0, \infty]$ \cite{Zhou1996Robust}. 

Interestingly, when $G$ is symmetric, i.e., $G(s)=G^T(s)$, the small gain condition is necessary in ensuring feedback stability for all symmetric norm bounded $H$, a subset of $\mathcal{B}[\gamma]$. Specifically, define  
\begin{equation*}
    \mathcal{B}_s[\gamma] = \{  H \in   \mathcal{B}[\gamma]: H(s) =H^T(s)   \}  , 
\end{equation*} 
which is a ball of symmetric systems.  

\begin{prop}             \label{prop:symmetric_small_gain_necessity} 
    Let $G \in \mathcal{RH}_\infty^{m \times m}$ be symmetric. Then $G\#H$ is stable for all $H \in \mathcal{B}_s[\gamma]$ if and only if 
    \begin{equation*}
        \overline{\sigma}(G(j\omega)) < |\gamma(\omega)|^{-1} 
    \end{equation*} 
    for all $\omega \in [0,\infty]$. 
\end{prop} 

Proposition \ref{prop:symmetric_small_gain_necessity} can be proved by applying Lemma \ref{lem:Autonne_Takagi_factorization} together with arguments analogous to those in \cite[Theorem 9.1]{Zhou1996Robust}. The details are omitted for brevity.   

In view of the necessity of the small gain condition against norm bounded uncertainties, a natural question arises: Is the small phase condition necessary against phase bounded uncertainties? This question, in its general form, appears rather challenging. Some preliminary explorations were made in \cite{Chen2024Phase,Mao2022Phases,Yang2023Small_phase_necessity,Yu2025Necessity_Small_Phase}. Nevertheless, in the next theorem we show that when $G$ is symmetric and frequency-wise quasi-sectorial, the small phase condition is necessary in ensuring feedback stability for all symmetric phase bounded $H$. This is in parallel to the result in Proposition \ref{prop:symmetric_small_gain_necessity}. Some extended discussions on the necessity of small phase condition for general asymmetric systems follow in the next section. 

Let $\alpha, \beta : \mathbb{R} \cup \{ \infty  \} \to \mathbb{R} $ be scalar functions satisfying $0 \leq \beta(\omega) - \alpha (\omega) < 2 \pi$. Define 
\begin{multline*} 
    \mathcal{C}[\alpha, \beta] \! = \! \{  H \in \mathcal{RH}_\infty^{m \times m}: H \text{ is frequency-wise semi-} \\ \text{ sectorial and }  \Phi(H(j\omega)) \! \subset \!  [\alpha(\omega), \beta(\omega)], \ \forall \omega   \in   [0, \infty]  \} . 
\end{multline*} 
and 
\begin{equation*} 
    \mathcal{C}_s [\alpha, \beta] = \{ H \in  \mathcal{C}[\alpha, \beta]: H(s) =H^T(s)   \} . 
\end{equation*}  

In view of the assumption that $H(0)$ is accretive, the scalar functions $\alpha(\omega)$ and $\beta(\omega)$ in $\mathcal{C}[\alpha,\beta]$ can be chosen such that $ [\alpha(0), \beta(0)] \subset [-\frac{\pi}{2},  \frac{\pi}{2}]$ without loss of generality. Moreover, $H(0)$ is real symmetric for all $H \in \mathcal{C}_s[\alpha, \beta]$, hence we can specialize $\alpha(0) = \beta(0) = 0$ in $\mathcal{C}_s[\alpha, \beta]$.

\begin{thm}         \label{thm:GH_both_symmetric} 
     Let $G \in \mathcal{R}^{m \times m}$ be symmetric and frequency-wise quasi-sectorial. Then $G\#H$ is stable for  all $H \in \mathcal{C}_s[\alpha, \beta]$ if and only if $G$ is stable and  
    \begin{equation}            \label{eq:phase_condition_necessity} 
        \Phi(G(j\omega)) \subset (-\pi - \alpha(\omega), \pi - \beta(\omega))  , 
    \end{equation} for all $\omega \in [0, \infty]$. 
\end{thm} 

\begin{pf}
    The sufficiency follows from Lemma \ref{lem:small_phase}. We show the necessity by contradiction. Suppose $G$ is not stable. Choose $H = 0$, then $H \in \mathcal{C}_s[\alpha, \beta]$ and $G\#H$ is not stable. Hence $G$ is stable.  Suppose there exists $\omega_0 \in [0, \infty]$ such that 
    \begin{equation}                \label{eq:contradiction_condition} 
        \Phi(G(j\omega_0)) \not\subset (-\pi - \alpha(\omega_0), \pi - \beta(\omega_0)) . 
    \end{equation} 
    Since $\alpha(0)= \beta(0)=0$ and by assumption, $G(0)$ is real positive semi-definite, clearly, $\omega_0 \neq 0$. 
    Then there are two cases. \\  
    \emph{Case 1. $\omega_0 = \infty$: } In this case, $G(\infty)$ is real symmetric and $G(\infty) \neq 0$. Then, there exists a nonzero eigenvalue $\lambda \in \mathbb{R}$ of $G(\infty)$ such that $\angle \lambda \notin (-\pi-\alpha(\infty), \pi - \beta(\infty))$. Choose 
    \begin{equation*}
        H(s) = h(s) I , 
    \end{equation*} 
    where $h(s) \in \mathcal{RH}_\infty$ is a scalar second order transfer function satisfying  $h \in \mathcal{C}_s[\alpha, \beta]$ and $h(\infty) = -\frac{1}{\lambda}$.  
    Then $H \in \mathcal{C}_s[\alpha, \beta]$ and $I+G(\infty)H(\infty)$ is singular.   \\ 
    \emph{Case 2. $\omega_0 \in (0, \infty)$: } In this case, $G(j\omega_0)$ is complex. Since $G(j\omega_0) = G^T(j\omega_0)$ and $G(j\omega_0)$ is quasi-sectorial, it follows from Corollary \ref{cor:complex_symmetric_quasi_matrix_decomposition} that there exists a nonsingular $T \in \mathbb{R}^{m \times m}$ such that 
    \begin{equation*}
        G(j\omega_0) = T^T  \mathrm{diag} \{ 0_{m-r},  e^{j\phi_1}, \dots, e^{j\phi_r} \}  T ,  
    \end{equation*}  
    with $r = \mathrm{rank}(G(j\omega_0))$ and  $\theta_0 + \frac{\pi}{2} > \phi_1 \geq \cdots \geq \phi_r > \theta_0 - \frac{\pi}{2}$ for some $\theta_0 \in \mathbb{R}$. In view of \eqref{eq:contradiction_condition}, we assume $\phi_1 \notin (-\pi-\alpha(\omega_0), \pi - \beta(\omega_0))$ without loss of generality. Then  choose 
    \begin{equation*} 
        H(s) = T^{-1}  \mathrm{diag} \{ 0_{m-r} , h(s) , 0_{r-1}  \} T^{-T} , 
    \end{equation*} 
    where $h(s) \in \mathcal{RH}_\infty$ is a scalar second order transfer function satisfying  $h \in \mathcal{C}_s[\alpha, \beta]$ and $h(j\omega_0) = e^{j(\pi - \phi_1)}$. Then $H \in \mathcal{C}_s[\alpha, \beta]$ and 
    \begin{align*}
        I+G(j\omega_0)H(j\omega_0) & = I + T^T \mathrm{diag}\{ 0_{m-r} , -1 , 0_{r-1} \} T^{-T}   \\ 
        & =  T^T \mathrm{diag} \{ 1_{m-r} , 0 , 1_{r-1} \} T^{-T} , 
    \end{align*} 
    implying $I+G(j\omega_0)H(j\omega_0)$ is singular. \\ 
    For both cases, singularity of $I+G(j\omega_0)H(j\omega_0)$ yields that $G\#H$ is not stable. This completes the proof. 
\end{pf}

\section{Discussions on Necessity of Small Phase Condition for Asymmetric Systems}
\label{section:discussions_general_LTI} 

We further discuss the necessity of small phase condition for general asymmetric systems. Consider the interconnection of $G\#H$ as shown in Fig.\ref{fig:feedback_system}, we are interested in the question as to whether the small phase condition \eqref{eq:phase_condition_necessity} is necessary in ensuring stability of $G\#H$ for all $H\in\mathcal{C}[\alpha,\beta]$. 

The following observation may give us some useful hints. For simplicity, suppose $G$ is stable and frequency-wise sectorial. Then for each frequency $\omega$, $G(j\omega)$ admits a symmetric polar decomposition \cite{Wang2020Phases}
\begin{equation*}
    G(j\omega) = P(j\omega) U(j\omega) P (j\omega) , 
\end{equation*} 
where $P(j\omega)$ is positive definite and $U(j\omega)$ is unitary. If the condition number of $P(j\omega)$ is close to $1$, then $G$ is close to being an inner system, i.e., $G^T(-s)G(s) = I$. The following proposition shows that for an inner and frequency-wise sectorial $G$, under some additional assumptions, the small phase condition is necessary for feedback stability against phase bounded uncertainties. 

\begin{prop}         \label{prop:inner_function_necessity} 
    Let $G \in \mathcal{RH}_\infty^{m \times m}$ be inner and frequency-wise sectorial satisfying 
    \begin{align} 
        \Phi(G(\infty)) & \subset (-\pi - \alpha(\infty), \pi - \beta(\infty)). 
        \label{eq:assumption_on_inner_infinity}
    \end{align} 
    Then $G\#H$ is stable for  all $H \in \mathcal{C}[\alpha, \beta]$ if and only if \eqref{eq:phase_condition_necessity} holds  
    for all $\omega \in [0, \infty]$. 
\end{prop} 

\begin{pf}
    The sufficiency follows from Lemma \ref{lem:small_phase}. For necessity, suppose by contradiction that there exists $\omega_0 \in [0, \infty]$ such that \eqref{eq:contradiction_condition} holds. \\ 
    In view of \eqref{eq:assumption_on_inner_infinity}, $\omega_0 \neq \infty$. Moreover, note that $[\alpha(0),\beta(0)] \subset [-\frac{\pi}{2}, \frac{\pi}{2}]$ and by assumption, $G(0)$ is accretive and sectorial, hence 
    \begin{equation*}
        \Phi(G(0)) \subset ( - \frac{\pi}{2}, \frac{\pi}{2} ) \subset (-\pi - \alpha(0), \pi - \beta(0)) , 
    \end{equation*} implying $\omega_0 \neq 0$. Therefore, the only possibility is that $\omega_0 \in (0, \infty)$. In this case,  
    $G(j\omega_0)$ is complex. Since $G$ is inner, there holds $G^T(-j\omega_0)G(j\omega_0) = I$. This implies $G(j\omega_0)$ is unitary and there exists a unitary matrix $U \in \mathbb{C}^{m \times m}$ such that 
    \begin{equation}                \label{eq:inner_unitary_decomposition} 
        G(j\omega_0) = U^H \mathrm{diag} \{  e^{j\phi_1}, \dots, e^{j\phi_m}   \} U,  
    \end{equation} 
    with $ \theta_0 + \frac{\pi}{2} > \phi_1 \geq \cdots \geq \phi_m > \theta_0 - \frac{\pi}{2}$ for some $\theta_0 \in \mathbb{R}$. Assume $\phi_1 \notin (-\pi-\alpha(\omega_0), \pi - \beta(\omega_0))$ without loss of generality. Then  choose 
    \begin{equation*}            
        H(s) = h(s) I , 
    \end{equation*} 
    where $h(s) \in \mathcal{RH}_\infty$ is a scalar second order transfer function satisfying  $h \in \mathcal{C}[\alpha, \beta]$ and $h(j\omega_0) = e^{j(\pi - \phi_1)}$. Then $H \in \mathcal{C}[\alpha, \beta]$ and  
    $I+G(j\omega_0)H(j\omega_0)$ is singular, which implies $G\#H$ is not stable. 
    This completes the proof. 
\end{pf} 

\begin{rem} 
    Note that without assumption \eqref{eq:assumption_on_inner_infinity}, Proposition \ref{prop:inner_function_necessity} no longer holds. This is because it would involve the necessity of small phase condition for real matrices, which is generally not true as noted in Remark \ref{rem:real_matrix_small_phase_not_hold}.   
\end{rem} 

When the condition number of $P(j\omega)$ is large, it reflects a significant distortion from an inner system. This deviation results in a considerable challenge. Specifically, suppose by contradiction that there exists $\omega_0 \in [0,\infty]$ such that \eqref{eq:contradiction_condition} holds. Decompose 
\begin{equation}       \label{eq:general_case_decomposition}
    G(j\omega_0) = T^H \mathrm{diag} \{ e^{j\phi_1}, \dots, e^{j\phi_m} \} T  , 
\end{equation} with  $\phi_1 \notin (-\pi-\alpha(\omega_0), \pi - \beta(\omega_0))$. Denote 
\begin{equation*} 
    H_0 = e^{j(\pi-\phi_1)} T^{-1}T^{-H} . 
\end{equation*} 
One can verify that $I+G(j\omega_0)H_0$ is singular. To show the necessity, we need to find a system $H\in \mathcal{C}[\alpha, \beta]$ that interpolates $H_0$ at $j\omega_0$. First, a real rational scalar function can be constructed to interpolate $e^{j(\pi-\phi_1)}$ at $j\omega_0$, which is straightforward. The challenge lies in constructing a real rational transfer function matrix with an arbitrarily small phase that interpolates the positive definite matrix $T^{-1}T^{-H}$ at $j\omega_0$. The interpolation is trivial if $T^{-1}T^{-H}$ is real-valued. However, it becomes substantially more difficult when $T^{-1}T^{-H}$ is complex-valued. We formalize this challenge into a general phase interpolation problem as follows, the solution of which remains open and needs deep investigation.

\textbf{Phase interpolation problem: }  
Given a pair $(j\omega_0,Z_0) $ with $\omega_0 \in [0,\infty], Z_0 \in \mathbb{C}^{m \times m}, Z_0 \geq 0$ and an arbitrarily small $\epsilon>0$, determine whether there exists $Z(s) \in \mathcal{RH}_\infty^{m \times m}$ satisfying $Z(j\omega_0) = Z_0$ and $\Phi_\infty(Z) \subset [-\epsilon,\epsilon]$.

\begin{rem} 
    When the systems are symmetric, the transformation matrix $T$ in \eqref{eq:general_case_decomposition} can be taken as real. Consequently, the matrix $Z_0$ in the above phase interpolation problem is real. In this case, $Z(s)=Z_0$ is a solution of the phase interpolation problem. 
\end{rem} 

\begin{rem}
    When $G$ is inner, the transformation matrix $T$ in \eqref{eq:general_case_decomposition} is unitary. As a result, the matrix $Z_0$ in the above phase interpolation problem is actually the identity matrix. In this case, $Z(s)=I$ is a solution of the phase interpolation problem. 
\end{rem}

\section{Conclusion} 
\label{section:conclusion}

In this paper, we have shown that the small phase condition is both sufficient and necessary for the stability of feedback interconnections of symmetric systems. The key lies in that for a complex symmetric and semi-sectorial matrix, the transformation matrix in the generalized sectorial decomposition can be take to be real. This result is significant and interesting in view of the wide applications of symmetric systems. Moreover, we explore the necessity of small phase condition for general asymmetric systems and formalize the main challenge as a phase interpolation problem, which remains open for future research.

\begin{ack}                               
The authors would like to thank Dr. Ding Zhang of Hong Kong University of Science and Technology, and Prof. Axel Ringh of Chalmers University of Technology and the University of Gothenburg for helpful discussions.                                           
\end{ack} 

\bibliographystyle{plain}        
\bibliography{autosam}           

\end{document}